\documentclass[twocolumn,superscriptaddress]{revtex4-1}
\usepackage[dvipdfmx]{graphicx}
\usepackage{listings}
\usepackage{color}
\usepackage{bm}
\usepackage{ascmac}
\usepackage{amsmath}
\usepackage{float}
\usepackage{ulem}
\begin{document}
\title{Ultrafast demagnetization in NiCo$_{2}$O$_{4}$ thin films probed by time-resolved microscopy}  
\author{Ryunosuke Takahashi}
\author{Yoshiki Tani}
\author{Hirotaka Abe}
\author{Minato Yamasaki}
\affiliation{Depertment of Material Science, Graduate School of Science, University of Hyogo, Ako, Hyogo 678-1297, Japan}
\author{Ikumi Suzuki}
\author{Daisuke Kan}
\author{Yuichi Shimakawa}
\affiliation{Institute for Chemical Research, Kyoto University, Uji, Kyoto 611-0011, Japan}
\author{Hiroki Wadati}
\affiliation{Depertment of Material Science, Graduate School of Science, University of Hyogo, Ako, Hyogo 678-1297, Japan}

\begin{abstract}
Using a time-resolved magneto-optical Kerr effect (TR-MOKE) microscope, we observed ultrafast demagnetization of inverse-spinel-type NiCo$_{2}$O$_{4}$  (NCO) epitaxial thin films of the inverse spinel type ferrimagnet NCO with perpendicular magnetic anisotropy. This microscope uses a pump-probe method, where the sample is pumped at 1030 nm, and magnetic domain images are acquired via MOKE microscopy at 515 nm (the second harmonic). We successfully observed the dynamics of the magnetic domain of the NCO thin film via laser irradiation, and obtained a demagnetization time constant of approximately 0.4 ps. This time constant was significantly smaller than the large time constants reported for other half-metallic oxides. These results, combined with the results of  our x-ray photoemission spectroscopy study, indicate that this NCO thin film is a ferrimagnetic metal whose electronic structure deviates from the theoretically predicted half-metallic one. 
\end{abstract}

\maketitle
Laser-induced reaction of magnetic materials is important in the development of magnetic data storage devices because laser pulse can control permanent magnetization reversal without contact. Photo-induced magnetism has been attracting considerable interest since Beaurepaire {\sl et al}. demonstrated an ultrafast demagnetization of a ferromagnetic Ni foil within 1 ps through the time-resolved magneto-optical Kerr effect (TR-MOKE) measurement with a pump-probe setup \cite{Beaurepaire1996}. Several challenges are encountered when observing ultrafast dynamics on magnetic materials with more than one magnetic element using TR-MOKE \cite{Kise2000,Guidoni2002,Ogasawara2005,Chinchetti2006,weber2011,Deb2018,Yamamoto2020} and time-resolved X-ray magnetic circular dichroism (XMCD) \cite{Radu2011,Higley2016,Takubo2017,Yamamoto2019}. Helicity-dependent magnetic domain images after laser irradiation were observed in ferrimagnetic the GdFeCo alloy using a MOKE (or Faraday) microscope  \cite{Stanciu2007,Ostler2012,TbCodomain2012}, whereas Lambert {\sl et al}. clearly captured the effect of laser-induced helicity-dependent demagnetization in ferromagnetic Pt/Co multilayer films via MOKE microscopy \cite{Lambert2014}. Under such circumstances, a time-resolved microscope emerged as a powerful tool to study the dynamics of magnetic domains and has been applied to study Gd$_{24}$Fe$_{66.5}$Co$_{9.5}$, the TbFeCo thin film, and the Ni$_{81}$Fe$_{19}$/Pt bilayer thin film \cite{Vahaplar2009,Ogasawara2009,Ganguly2015}. These studies revealed that the demagnetization time constant $\tau_m$ strongly depends on the type of magnetic material; according to the phenomenological three-temperature model, $\tau_m$ was determined as the strength of coupling between electrons and spins \cite{Muller2009,Kirilyuk2010}. The mechanism of the electron-spin coupling is still under intense debate. 

$\tau_m$ depends on the electron and spin states of the materials (specifically, the insulator, metal, or half metal) \cite{Kirilyuk2010}. For instance, Tsuyama {\sl et al}. observed changes in  $\tau_m$ from $\sim$ 150 ps (slow) to $<$ 70 ps (fast), which can be attributed to the transition into a metallic state induced by laser excitation in ferromagnetic insulating BaFeO$_{3}$ thin films \cite{Tsuyama2016}. M\"{u}ller {\sl et al}. reported that laser-induced demagnetization in half-metallic oxides has longer $\tau_m$ than Ni in terms of speed owing to the weakness of electron–spin interaction in one spin channel \cite{Muller2009}. Half-metallic oxides, CrO$_{2}$, La$_{0.66}$Sr$_{0.33}$MnO$_{3}$, and Fe$_{3}$O$_{4}$, have been investigated, and $\tau_m$ of these materials was determined as $\tau_{m}$ = 200 – 300 ps, $\tau_{m} >$ 1 ns, and 400–600 ps, respectively, in sharp contrast to ultrafast demagnetization in Ni within 1 ps \cite{Muller2009}.  

NiCo$_{\bm{2}}$O$_{\bm{4}}$ (NCO) with an inverse-spinel-type crystal structure is a ferrimagnet whose transition temperature is higher than 400 K \cite{Knop1968}, and the half-metallic electronic state was predicted based on the first principle calculation result \cite{Ndione2014}. Co occupies the T$_{d}$ and O$_{h}$ sites, and Ni occupies the O$_{h}$ site. The valence states are close to Co$^{2+}$ (O$_{h}$ site), Co$^{3+}$ (T$_{d}$ site), and Ni$^{2+\delta}$ (O$_{h}$ site), as determined by X-ray absorption spectroscopy and XMCD measurement \cite{Kan2020prb}. NCO thin films exhibit perpendicular magnetic anisotropy (PMA), attracting attention as a spintronic material. The spin direction of O$_{h}$-site Co and T$_{d}$-site Ni is ferrimagnetically coupled \cite{Bitla2015,Kan2020prb}, and the saturated magnetization has been determined as $\sim$2$\mu_{B}$ \cite{Kan2020}. When NCO is epitaxially grown, the cation occupations deviate from the ideal ones depending on growth conditions. For NCO epitaxial films developed by pulsed laser deposition, the T$_{d}$-site Co is partially replaced with the O$_{h}$-site Ni, influencing magnetic and transport properties.  By depositing NCO at relatively high oxygen pressure (e.g., 100 mTorr), they have been shown to produce epitaxial films whose cation occupation is close to the stoichiometric \cite{Kan2020,Shen2020}. 

We study the TR-MOKE of laser irradiated magnetic domains in NCO thin films at room temperature. Based on the TR-MOKE microscopy measurements, we have revealed that the value of $\tau_{m}$ $\sim$ 0.4 ps is independent of the fluence of pump pulse. $\tau_{m}$ of $\sim$ 0.4 ps is much shorter than that of other half-metallic oxides ($\tau_{m} >$ 100 ps). Considering the metallicity based on the X-ray photoemission spectroscopy (XPS) measurement, this result implies that the spin polarization $|P|$ of the NCO film is $\sim$0.7, lower than the value expected from theoretically predicted half-metallic band structure \cite{Shen2020}.  

Epitaxial NCO thin films with a thickness of 30 nm were grown on MgAl$_{\bm{2}}$O$_{\bm{4}}$ (100) substrates using pulsed laser deposition. During the film deposition, the substrate temperature and oxygen pressure were 315$^\circ$C and 100 mTorr, respectively. The NCO samples are described in Refs. \cite{Kan2020,Shen2020}.
\begin{figure}[H]
\centering
\includegraphics[scale=0.46]{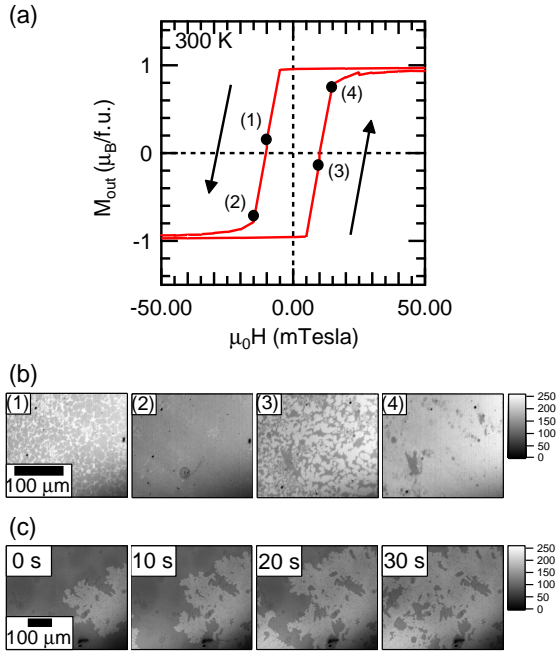}
\caption{(a) Magnetization curve of the sample as a function of the field out of plane. (b) Field-dependent magnetic domain images of an NCO thin film. (c) Slow domain wall motion with external magnetic field $B$ = 6.5 mT. The scale bar in (b, c) is 100 $\mu$m. }
\label{figFD}
\end{figure}
\vspace{-0.4cm}
First, we have measured the field-dependent magnetic domain for NCO thin films based on MOKE microscopy using white LED as a light source with K\"{o}hler illumination method. The magnitude of the external magnetic field was controlled using a Nd magnet ({\sl B} = 500 mT). When a magnetic field is applied, we observed a domain wall movement causing magnetization reversal. Figure \ref{figFD} shows the $M$-$H$ curve and magnetic domain images. The response of magnetic domains corresponds to the measured magnetization curve of the previous study \cite{Kan2020} (Fig. \ref{figFD} (a)). An NCO thin film has a bubble domain structure, and its sizes change with the application of the external magnetic field (Fig. \ref{figFD} (b)). The slow domain wall motion with the external magnetic field $B$ = 6.5 mT was observed, as shown in Fig. \ref{figFD} (c). The domain walls crept in the order of $\sim1 \mu$m/s. This motion was so slow that we waited until the domain walls stopped moving and then captured the images (Fig. \ref{figFD} (b)).

To investigate the electronic state near E$_{F}$, we performed the XPS measurement of an NCO thin film. The monochromatized Al $K\alpha$ ($h\nu$ = 1486.6 eV) line was used to irradiate the NCO thin films. Binding energies were calibrated using the E$_{F}$ position of Ag, which is in electrical contact with the sample. The total energy resolution 
\begin{figure}[H]
\centering
\includegraphics[width = 7.9cm]{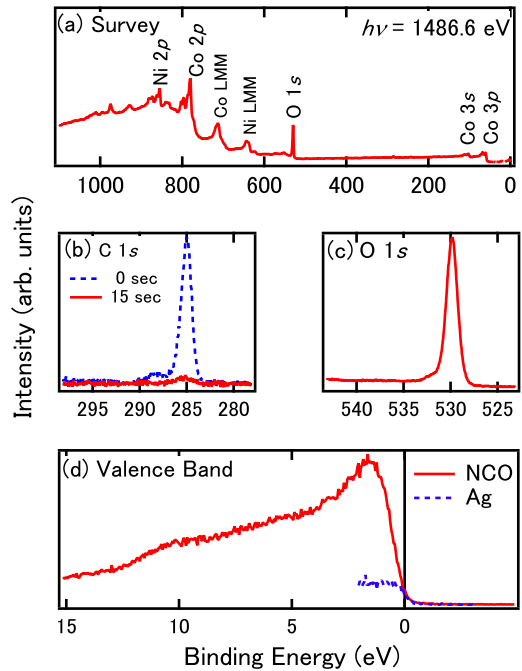}
\caption{XPS spectra of (a) survey, (b) C 1{\sl s}, (c) O 1{\sl s}, and (d) valence band of an NCO thin film.}
\label{figxps}
\end{figure}
\noindent was $\Delta E$ = 530 meV. Ar sputtering was carried out (1 kV, 2 mm$\times$2 mm, 15 s) to obtain clean surfaces. Figure \ref{figxps} shows the XPS spectra of an NCO thin film after Ar sputtering. The cleanliness of the sample surface was confirmed by the disappearance of the C 1{\sl s} peak after Ar sputtering (Fig. \ref{figxps} (a, b)), and by the absence of the contamination signal on the higher-binding-energy side of the O 1{\sl s} peak (Fig. \ref{figxps} (c)). The valence-band spectrum in Fig. \ref{figxps} (d) shows a finite electronic state 
\begin{figure}[H]
\centering
\includegraphics[scale=0.37]{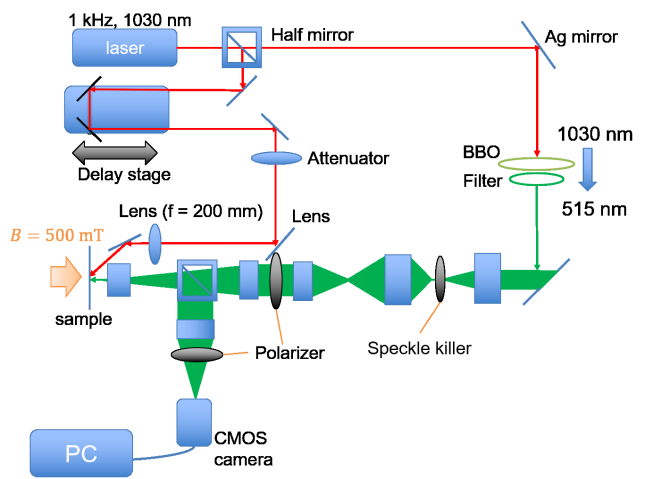}
\caption{Diagram of the TR-MOKE microscope}
\label{figsetup}
\end{figure}
at E$_{F}$, which was confirmed by the overlapping of the Ag spectrum. This result of E$_{F}$ crossing indicates the metallic nature of the sample.

Next, let us discuss the results of the TR-MOKE microscopy measurement of the NCO thin films. Figure \ref{figsetup} shows the experimental setup of TR-MOKE microscopy with the pump-probe method. Femtosecond Yb:KGW laser Pharos ($\lambda$ = 1030 nm ($\sim$ 1.2 eV), 1 kHz, FWHM $\sim$ 0.2 ps) was used as a light source. The NCO thin films were excited by pump pulses ($\lambda$ = 1030 nm) and observed by probe pulses, which had half wavelength ($\lambda$ = 515 nm) based on the second harmonic generation of the nonlinear crystal beta barium borate ($\beta$-BaB$_{\bm{2}}$O$_{\bm{4}}$). Using an optical color filter, the infrared beam is filtered after the BBO crystal. We focused on the pump pulses with a lens of focal length 200 mm, and the diameter of the pump pulse was 240 $\mu$m. The magnetization of the sample was saturated using the Nd magnet ({\sl B} = 500 mT) backward during the TR-MOKE microscopy measurement. We adopted a K\"{o}hler illumination of the sample and reduced the speckle pattern of the laser using a speckle killer.  
\begin{figure}[H]
\centering
\includegraphics[scale = 0.5]{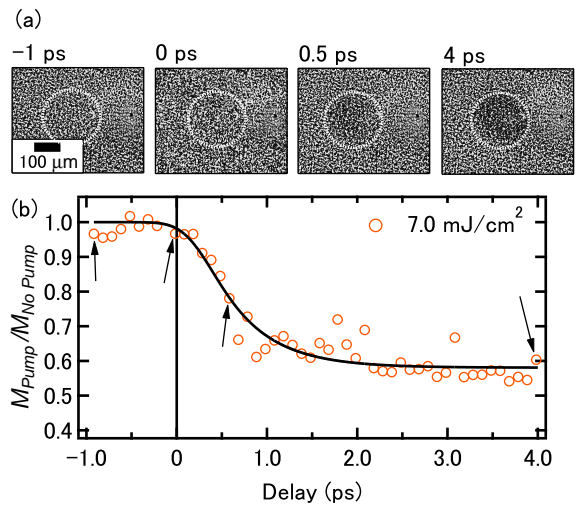}
\caption{TR-MOKE microscopy measurement of an NCO thin film. The scale bar is 100 $\mu$m. Panel (a) shows magnetic domain images at delay points under laser irradiation. The dotted circles in panel (a) denote the pump-irradiated areas. Panel (b) shows the delay scan when the pump fluence is 7.0 mJ/cm$^2$.}
\label{fig7mJ}
\end{figure}
 Figure \ref{fig7mJ} shows the dynamics of laser-induced demagnetization when the pump fluence is 7.0 mJ/cm$^2$. Panel (a) shows the magnetic domain images at delay $= -1, 0, 0.5,$ and $4$ ps, and the dotted circles denote the laser-irradiated areas. We obtained the domain images that were captured at each delay point, and plotted the average contrast value of the pumped area, as shown by circles in (a). Figure \ref{fig7mJ} (b) shows the delay scan of $M$$_{pump}$ obtained from images in (a). Solid line in (b) was the fitting curve obtained using the following equation:
\begin{equation}
\frac{M_{Pump}}{M_{No\,Pump}}(t) = 1+\theta(t) \times A(e^{-\frac{t}{\tau_{m}}}-1),
\end{equation}
where $\theta (t)$ is the Heviside step function. We determined the $\tau_{m}$ and demagnetization magnitude {\sl A}, considering the fitting function (Eq. (1)) convoluting with a Gaussian function of FWHM = 0.3 ps by the auto-correlation of the laser pulse. We obtained the time constant $\tau_{m}$ $\sim$ 0.6 ps, and $A$ = 0.4 was determined in this case at a fluence of 7.0 mJ/cm$^2$.
\begin{figure}[H]
\centering
\includegraphics[width = \linewidth]{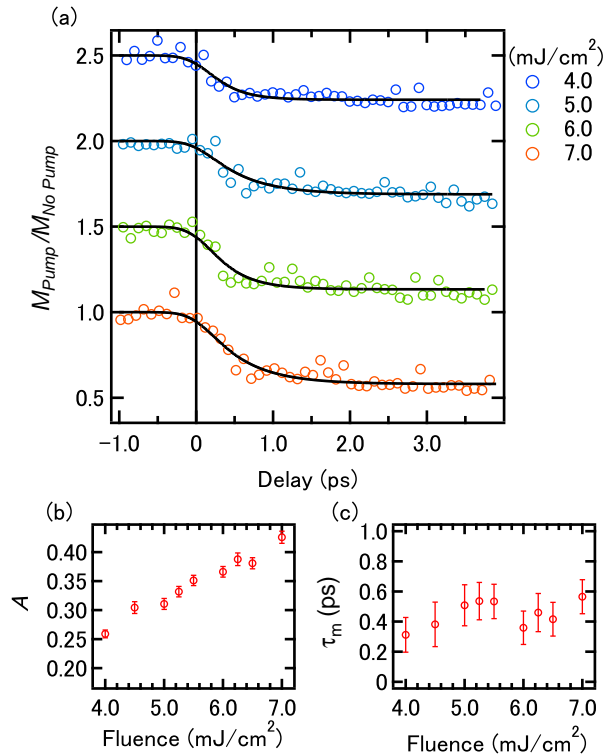}
\caption{Pump-fluence dependence. (a) Delay scans, (b) magnitude of change, and (c) time constant $\tau_{m}$ of an NCO thin film.}
\label{figTRMOKE}
\end{figure}
To investigate the effect of pump pulse fluence on $\tau_{m}$ and {\sl A}, we measured the fluence-dependent magnetization dynamics (Fig. \ref{figTRMOKE}). Figure \ref{figTRMOKE} (a) shows the results of delay scan pertaining to various fluences. This result is systematically fitted by Eq. (1), and the values of {\sl A} and $\tau_{m}$ are shown in panels (b) and (c), respectively. {\sl A} increases with the increase in fluence, and the demagnetization strength becomes stronger. However, the $\tau_{m}$ of 0.4 ps is almost independent on the pump fluence, demonstrating the intrinsic demagnetization time constant of the NCO thin films. The time constant of half metals shows a power dependence on spin polarization $P$ \cite{Muller2009}. The $P$ value of perpendicularly magnetized NCO thin films was $\sim$0.7, as determined using tunnel magnetoresistance \cite{Shen2020apl}.

Next, we discuss the relatively fast time constant of an NCO thin film in contrast to other half-metallic oxides. Based on the relationship between $P$ and $\tau_{m}$ in Ref. \cite{Muller2009}, $\tau_{m}$ $\sim$ 0.4 ps provides $P$ $\sim$ 0.7. In the theoretically predicted half-metallic band structure in NCO \cite{Shen2020}, the density of states at E$_{F}$ comprises the minority-spin subband, while the majority-spin subband has an energy gap (the half-metallic energy gap) of $\sim$ 0.5 eV \cite{Ndione2014}, which is close to that of Heusler alloy  Co$_{2}$MnSi (0.64 eV) \cite{Muller2009}. This energy gap is much smaller than those of other half-metals that exhibit relatively slow demagnetization behavior. Moreover, we indicate that the magnetic and transport properties of NCO films are significantly affected by deviations in the cation occupation of the stoichiometric one \cite{Shen2020}, implying that the cation occupation plays a key role in the electronic structures of NCO films. In fact, even for NCO films whose cation occupation is close to stoichiometric, some amounts of Ni are still accommodated in the T$_{d}$ site \cite{Shen2020}. Such deviations in the cation occupation modify the half-metallic band structure and reduce spin polarization. Furthermore, NCO films have sufficient PMA ($K$$_{1}$ $\sim$ $1 \times 10^5$ J/m$^3$) based on the spin-orbit interaction, which might contribute to the fast demagnetization process \cite{koizumi2021spin}.

In summary, we studied the TR-MOKE microscope of inverse-spinel-type half-metallic-oxide NCO epitaxial thin films with PMA. This microscope used a pump-probe method using 1030-nm pulse laser as a pump and magnetic domain images were captured using MOKE microscopy with 515-nm probe. We succeeded in capturing the clear change in magnetic domains caused by pump laser irradiation. We observed laser-induced demagnetization at $\tau_{m}$ $\sim$ 0.4 ps. This ultrafast timescale is consistent with the $P$ values of the NCO thin films determined by tunnel magnetoresistance \cite{Shen2020apl}. 

We would like to thank T. Ogasawara for informative discussions.~This study was supported by Japan Society for the Promotion of Science (JSPS) Core-to-Core Program (A. Advanced Research Networks) under the grant for the Integrated Research Consortium on Chemical Sciences and the International Collaborative Research Program of Institute for Chemical Research in Kyoto University, Ministry of Education, Culture, Sports, Science and Technology (MEXT), Japan.
This work was supported by the Yamada Science Foundation, JSPS KAKENHI under Grant Nos.~19H01816, 19H05816, 19H05823, 19H05824, and 21H01810 KAKENHI and
MEXT Quantum Leap Flagship Program (MEXT Q-LEAP) under Grant No. JPMXS0118068681.

The data that support the findings of this study can be obtained
from the corresponding author upon a reasonable request.

\bibliography{Mycollection2}

\end{document}